\documentclass[10pt,conference]{IEEEtran}
\IEEEoverridecommandlockouts
\usepackage{amsmath,amssymb,amsfonts}                           
\usepackage{hyperref}                                           
\usepackage[caption=false]{subfig}                              
\usepackage[linesnumbered,algoruled,boxed,lined]{algorithm2e}   
\usepackage[dvipsnames]{xcolor}
\usepackage{tikz}                                               
\usetikzlibrary{quantikz2}                                      
\usepackage{blindtext}
\usepackage{cleveref}
\usepackage{graphicx}
\usepackage{orcidlink}
\usepackage{etoolbox}
\makeatletter
\patchcmd{\@makecaption}
  {\scshape}
  {}
  {}
  {}
\makeatletter
\patchcmd{\@makecaption}
  {\\}
  {.\ }
  {}
  {}
\makeatother
\def\tablename{Tab.}

\usepackage{bbold}
\usepackage{multirow}
\usepackage{upgreek}
\usepackage{cite}

\DeclareMathOperator*{\argmin}{arg\,min}

\newcommand{\erdosrenyi}{Erd\H{o}s-R\'{e}nyi}

\def\linkurl#1{\url{#1}}

\begin{document}

\title{Towards Robust Benchmarking of Quantum Optimization Algorithms}


\makeatletter
\newcommand{\linebreakand}{%
  \end{@IEEEauthorhalign}
  \hfill\mbox{}\par
  \mbox{}\hfill\begin{@IEEEauthorhalign}
}
\makeatother

\author{
\IEEEauthorblockN{%
David Bucher\IEEEauthorrefmark{1}\textsuperscript{$\orcidlink{0009-0002-0764-9606}$},
Nico Kraus\IEEEauthorrefmark{1},
Jonas Blenninger\IEEEauthorrefmark{1},
Michael Lachner\IEEEauthorrefmark{1}\textsuperscript{$\orcidlink{0009-0008-6874-8329}$},
Jonas Stein\IEEEauthorrefmark{1}\IEEEauthorrefmark{2}\textsuperscript{$\orcidlink{0000-0001-5727-9151}$}
and Claudia Linnhoff-Popien\IEEEauthorrefmark{2}\textsuperscript{$\orcidlink{0000-0001-6284-9286}$}}
\IEEEauthorblockA{\IEEEauthorrefmark{1}\textit{Aqarios GmbH, Munich, Germany}}
\IEEEauthorblockA{\IEEEauthorrefmark{2}\textit{LMU Munich, Institute for Computer Science, Munich, Germany}}
\IEEEauthorblockA{\{david.bucher, nico.kraus, jonas.blenninger, michael.lachner, jonas.stein\}@aqarios.com}
}

\maketitle

\bstctlcite{BSTcontrol}

\begin{abstract}
Benchmarking the performance of quantum optimization algorithms is crucial for identifying utility for industry-relevant use cases. Benchmarking processes vary between optimization applications and depend on user-specified goals. The heuristic nature of quantum algorithms poses challenges, especially when comparing to classical counterparts.
A key problem in existing benchmarking frameworks is the lack of equal effort in optimizing for the best quantum and, respectively, classical approaches.
This paper presents a comprehensive set of guidelines comprising universal steps towards fair benchmarks.
We discuss (1) application-specific algorithm choice, ensuring every solver is provided with the most fitting mathematical formulation of a problem; (2) the selection of benchmark data, including hard instances and real-world samples; (3) the choice of a suitable holistic figure of merit, like time-to-solution or solution quality within time constraints; and (4) equitable hyperparameter training to eliminate bias towards a particular method.
The proposed guidelines are tested across three benchmarking scenarios, utilizing the Max-Cut (MC) and Travelling Salesperson Problem (TSP). The benchmarks employ classical mathematical algorithms, such as Branch-and-Cut (BNC) solvers, classical heuristics, Quantum Annealing (QA), and the Quantum Approximate Optimization Algorithm (QAOA).
%

\end{abstract}

\section{Introduction}
\label{sec:introduction}
The ever-growing interest in applying quantum optimization algorithms to industry-relevant Combinatorial Optimization Problems (COPs) requires a robust benchmarking framework. 
While a substantial body of literature exists, proposing benchmarks for quantum and classical algorithms~\cite{kuramata2022,finzgar2022,blekos2023,lubinski2024optimization,tasseff2022emerging,king2015, pelofske2023,Au2023,crooks2019performance}, some fall short of being applicable to comparing real-world applications \emph{fairly}. Despite real-world instances sometimes being considered~\cite{kuramata2022, finzgar2022}, comparisons predominantly rely on randomly generated problems~\cite{blekos2023, lubinski2024optimization} or even crafted instances suiting the quantum hardware~\cite{tasseff2022emerging, king2015, pelofske2023}.
Besides benchmarks focusing on quantum-only comparisons, those also investigating classical methods sometimes lack the proper selection of use-case-optimized classical algorithms~\cite{Au2023, tasseff2022emerging} but instead rely on a small set of standard methods~\cite{finzgar2022, crooks2019performance}, whereas quantum methods are highly optimized for. This leads to suboptimal results for the classical approaches. In addition, the problem formulation generally has an enormous impact on the solution quality for both quantum~\cite{Zielinski2023} and classical algorithms~\cite{Taveres2021,tasseff2022emerging,seker2022} and is not necessarily optimal for the solver at hand.

\definecolor{ceruleanblue}{RGB}{42, 82, 190}
\definecolor{powederblue}{RGB}{176, 224, 230}
\definecolor{teal}{RGB}{0, 128, 128}
\definecolor{lavender}{RGB}{181, 126, 220}
\definecolor{mintgreen}{RGB}{152, 255, 152}
\definecolor{softslate}{RGB}{112, 128, 144}

\colorlet{copapplication}{CadetBlue!20}
\colorlet{fom}{Orchid!20}
\colorlet{algorithms}{Thistle!20}
\colorlet{dataset}{Salmon!20}
\colorlet{hyperparameter}{Apricot!20}
\colorlet{benchmark}{Goldenrod!20}

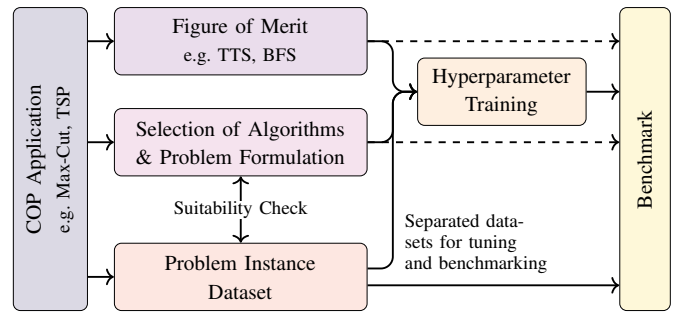
\begin{figure}
    \centering
    \resizebox{\columnwidth}{!}{%
    \begin{tikzpicture}[mbox/.style={rounded corners=3pt}, marrow/.style={rounded corners=5pt, ->, thick}]
        \node[draw, minimum width=1.1cm, minimum height=4.5cm, anchor=south west, inner sep=0, fill=copapplication, mbox] (start) at (0,0) {};
        \node[rotate=90, align=center] at (0.55, 2.25) {\small COP Application\\\footnotesize e.g. Max-Cut, TSP};
 
        \node[draw, minimum width=3.75cm, minimum height=1cm, anchor=south west, inner sep=0, fill=fom, mbox] (merit) at (1.5,3.5) {};
        \node[align=center] at (3.375, 4) {\small Figure of Merit\\ \footnotesize e.g. TTS, BFS};
 
        \node[draw, minimum width=3.75cm, minimum height=1cm, anchor=south west, inner sep=0, fill=algorithms, mbox] (sel) at (1.5,2) {};
        \node[align=center] at (3.375, 2.5) {\small Selection of Algorithms \\ \small\& Problem Formulation};

        \node[draw, minimum width=3.75cm, minimum height=1cm, anchor=south west, inner sep=0, fill=dataset, mbox] (dset) at (1.5,0) {};
        \node[align=center] at (3.375, 0.5) {\small Problem Instance \\ \small Dataset};
        
        \draw[marrow] ([yshift=1.75cm]start.east) -- (merit.west);
        \draw[marrow] ([yshift=+0.25cm]start.east) -- (sel.west);
        \draw[marrow] ([yshift=-1.75cm]start.east) -- (dset.west);
 
        \draw[<->, thick] (dset.north) -- node[fill=white, inner sep=0pt]{\footnotesize Suitability Check} (sel.south);
 
        \node[draw, minimum width=2.5cm, minimum height=1cm, anchor=south west, inner sep=0, fill=hyperparameter, mbox] (hopt) at (6,2.75) {};
        \node[align=center] at (7.25, 3.25) {\small Hyperparameter \\ \small Training};
        
        \draw[marrow] (sel.east) -- ++(0.375,0) |- (hopt.west);
        \draw[marrow] ([yshift=0.125cm]dset.east) -- ++(0.375,0) |- (hopt.west);
        \draw[marrow] (merit.east) -- ++(0.375,0) |- (hopt.west);
        
        \node[draw, minimum width=0.75cm, minimum height=4.5cm, anchor=south west, inner sep=0, fill=benchmark, mbox] (bench) at (9,0) {};
        \node[rotate=90] at (9.375, 2.25) {\small Benchmark};
        
        \draw[marrow, dashed] (merit.east) -- ([yshift=1.75cm]bench.west);
        \draw[marrow, dashed] (sel.east) -- ([yshift=0.25cm]bench.west);
        \draw[marrow] ([yshift=-0.125cm]dset.east) -- ([yshift=-1.875cm]bench.west);
        
        \draw[marrow] (hopt.east) -- ([yshift=1cm]bench.west);
        
        \node[align=left] at (6.875, 1.0) {\footnotesize Separated data-\\[-0.3em] \footnotesize sets for tuning\\[-0.3em]\footnotesize and benchmarking};

    \end{tikzpicture}
    }
    \caption{Overview of the benchmarking guidelines devised throughout this paper.}
    \label{fig:overview}
\end{figure}

As benchmarking depends on a wide variety of requirements and criteria defined by the investigated problem, the considered solvers, the benchmarking objective, and the dataset at hand, developing an application-agnostic approach is hardly possible. Instead, this paper proposes a set of guidelines and steps that need to be followed in order to ensure a \emph{fair}---and thus meaningful---comparison given a specific use-case. By targeting industrial application scenarios, our approach consequently contributes to establishing best practices for investigating practically relevant quantum utility. As outlined in Fig.~\ref{fig:overview}, we concretize the following key points:

Central to benchmarking is the choice of state-of-the-art algorithms for the problem, both quantum and classical. For that, one should emphasize that the mathematical problem formulation plays a crucial role in the solver's performance, e.g., a classical BNC solver~\cite{BnC} typically shines in performance when posed with a Mixed-Integer Linear Program (MILP)~\cite{bhatti2012practical} formulation of a constrained problem instead of the Quadratic Unconstrained Optimization Problem (QUBO) reformulation with penalty terms, required for quantum solvers.

The benchmarking dataset should ideally consist of a diverse and real-world-inspired collection of problem instances~\cite{abbas2023quantum}. However, suitability considerations may require randomly generated instances, e.g., if the problem size is limited due to simulation or device restrictions~\cite{preskill2018quantum}.

To quantify solver performance, a holistic figure of merit is required that includes the trade-off between solution quality and runtime that is inherently present in heuristic algorithms~\cite{lubinski2024optimization}. Such a figure of merit can be the Best Solution Found (BSF)~\cite{Oshiyama2022} within given time bounds or the Time-To-Solution (TTS)~\cite{albash2018, lubinski2024optimization} and can be utilized for fair hyperparameter optimization.
Alternatively, the Pareto front of the joint solution quality and runtime combination can also be a meaningful comparison tool.

We demonstrate the practical applicability of our guidelines on three optimization scenarios, consisting of two Max-Cut (MC) problems at different sizes and the Travelling Salesperson Problem (TSP). In the last case, we focus on the comparison of different designs of quantum algorithms.


The remainder of this paper is structured as follows: First, we outline background concepts for solving COPs in Sec.~\ref{sec:background}. Then, Sec.~\ref{sec:methodology} introduces the guidelines in detail, and Sec.~\ref{sec:examples} shows exemplary application of those. Finally, we conclude with Sec.~\ref{sec:conclusion}.

\section{Background}
\label{sec:background}
In combinatorial optimization, one generally seeks an optimal element $x^* \in \mathcal{X}$ in a finite set of possible solutions, i.e., w.l.o.g. $x^* = \argmin_{x\in \mathcal{X}} C(x)$ for some objective function $C: \mathcal{X} \rightarrow \mathbb{R}$~\cite{korte2011combinatorial}. As the set of possible solutions typically scales exponentially in the problem size, and as many of the problems of interest are NP-hard, one is limited to finding approximately optimal solutions in practice. As a consequence of the no-free-lunch theorem, which essentially states that no single algorithm can perform best on all possible problem instances, many different heuristics have been developed~\cite{585893}. To gain some more insights into practically employed approaches, we provide a list of selected classical and quantum state-of-the-art solvers.

\textbf{Tabu Search} (TS)~\cite{GLOVER1986533,glover1989tabu,glover1990tabu,palubeckis2004}: Given an initial solution $x\in \mathcal{X}$, tabu search iteratively explores the search space $\mathcal{X}$ by moving to the best neighboring solution $x\leftarrow x' \in \mathcal{N}(x)\setminus \mathcal{T}$, where $\mathcal{N}(x)\subseteq \mathcal{X}$ denotes the set of all possible neighbors of $x$, which is typically defined based on expert knowledge in the domain of the optimization problem. The set $\mathcal{T}\subseteq \mathcal{X}$ is called \emph{tabu list} and contains a fixed maximal number of solutions that must not be visited again. It is filled first-in-first-out by adding the current candidate solution $x'$ after each iteration. This eventually allows for escaping local minima. Finally, TS keeps track of the best solution found and returns it after reaching a user-specified maximum number of iterations. The heuristic is commonly repeated for many initial conditions. We use the TS implementation of~\cite{stevanovic2023}.

\textbf{Simulated Annealing} (SA)~\cite{kirkpatrick1983optimization}: Like TS, Simulated Annealing performs a local search starting from a given initial solution $x\in \mathcal{X}$. Instead of selecting the best neighboring solution as in TS, SA iteratively explores a randomly selected neighboring solution $x'\in \mathcal{N}(x)$. This candidate solution is always accepted (i.e., $x\leftarrow x'$) if it has a better objective value than the current solution, but only with probability $P(x,x',T_i)$ if its objective value is worse. Here, $P(x,x',T) = e^{-\left|C(x)-C(x')\right|/kT}$, where $k$ denotes a user-specified constant (historically the Boltzmann constant) and $T_i$ the \emph{temperature}, which is updated in each iteration $i$, based on a chosen \emph{annealing schedule}. This makes SA a Markov chain Monte Carlo method. A popular choice is the \emph{geometric} annealing schedule, which decreases the temperature $T_i$ exponentially via $T_i\leftarrow \alpha T_{i-1}$, where $T_0$ denotes the initial temperature and $\alpha \in (0,1)$ the decay hyperparameter~\cite{Yaghout_Nourani_1998}. Finally, SA yields the best solution found after a specified maximal number of iterations. We use the SA implementation of~\cite{stevanovic2023}.

\textbf{Commercial Solvers} (CPLEX~\cite{CPLEX} and Gurobi~\cite{gurobi}): The state-of-the-art in solving mixed integer optimization problems is predominantly found in proprietary, commercial products such as CPLEX and Gurobi. Their core solvers are based on the branch-and-cut (BNC) algorithm~\cite{BnC}, which starts by executing the cutting plane method~\cite{Cpm}, followed by the branch-and-bound algorithm~\cite{Gamrath2020}. In the cutting plane method, the problem is first relaxed to a continuous domain and then solved with a Linear Programming (LP) solver (e.g., the simplex algorithm~\cite{dantzig1951maximization} or the interior point method~\cite{interiorpointmethod}). Finally, through the iterative addition of heuristic constraints, the solution space gets increasingly restricted until (ideally) an integer solution is found, i.e., without cutting away solutions from the original domain space. If the cutting plane method is unable to find an integer solution, the branch-and-bound algorithm is started subsequently. Here, the domain space is heuristically split into subspaces, for which lower bounds on the best possible solution are identified using an LP solver. Using this branching iteratively, a tree for searching the solution space systematically is constructed. The gap between the lower bound and the best solution is continuously closed until optimality is proven. Compared to other open source solvers like GLPK~\cite{GLPK}, SCIP~\cite{SCIP}, lp\_solve~\cite{LP_solve}, or COIN-OR~\cite{COIN-OR}, CPLEX and Gurobi have shown to yield better results on average~\cite{mittelmann, sun2024}, which can likely be attributed to their highly optimized (problem-aware) heuristics for selecting the cutting planes and the branching criteria.

\textbf{Quantum Annealing} (QA)~\cite{PhysRevE.58.5355} is built upon the adiabatic principle of quantum mechanics.
In essence, the adiabatic theorem states that a quantum mechanical system remains in its lowest energy state when the changes applied to it are carried out adiabatically, i.e., slowly enough~\cite{Born1928}. When applied to the transverse field Ising model, this allows for solving a large class of classical optimization problems~\cite{PhysRevE.58.5355}.

More concretely, the cost landscape of any Quadratic Unconstrained Binary Optimization (QUBO) problem
\begin{align}
    \argmin_{x\in\left\lbrace0,1\right\rbrace^n} \sum_{i,j} Q_{ij}x_i x_j
\end{align}
with $Q\in\mathbb{R}^{n\times n}$, can be straightforwardly mapped onto the energy landscape of an Ising model
\begin{align}
    \hat{H}_C=\sum_{i=1}^n h_i\hat{\sigma}_{i}^z + \sum_{i,j} J_{ij}\hat{\sigma}_{i}^z\hat{\sigma}_{j}^z
\end{align}
by identifying $1 - 2x_i \leftrightarrow \hat{\sigma}_{i}^{z}$, where $\hat{\sigma}_{i}^{z}$ denotes the Pauli-$z$ matrix acting on the $i$-th qubit.
Initializing a system in the equal superposition $\ket{+}^{\otimes n}$, the ground state of the transverse field mixer Hamiltonian $\hat{H}_M = - \sum_i \hat{\sigma}^x_{i}$, and adiabatically shifting the system towards $\hat{H}_C$, allows for solving arbitrary QUBO problems. In reality, the adiabatic transition is generally disallowed because of short-lived noisy qubits. However, a faster-than-adiabatic transition (i.e., QA) prepares final quantum states with a high probability of measuring near-optimal solutions. Therefore, similar to TS and SA, one needs to repeat the process and sample from the output distribution.

\textbf{Quantum Approximate Optimization Algorithm} (QAOA)~\cite{farhi2014quantum}: While Quantum Annealers can currently tackle the largest problem instances as they offer the highest number of (practically usable) physical qubits, their analog form of computation makes them heavily susceptible to hardware noise. A straightforward approach to fix this issue is simulating the process of Quantum Annealing on a gate-based quantum computer, which allows for error correction. In essence, this demands an algorithm for simulating the time-dependent Hamiltonian
\begin{align}
    \hat{H}(s)=\left(1 - s\right)\hat{H}_M + s\hat{H}_C
\end{align}
where $s$ monotonically transitions from 0 to 1. By discretizing this continuous time evolution in $p\in\mathbb{N}$ rounds, one can derive the following quantum operation, i.e., the QAOA
\begin{align}
    U\left(\beta, \gamma\right) = U_M(\beta_p) U_C(\gamma_p) \ldots U_M(\beta_1) U_C(\gamma_1),
\end{align}
where $U_M(\beta_i) = e^{-i\beta_i \hat{H}_{M}}$, $U_C(\gamma_i) = e^{-i\gamma_i \hat{H}_{C}}$. 
$U\left(\beta, \gamma\right)$ approaches adiabatic evolution for $p\rightarrow \infty$, and constant speed, i.e, $\beta_i = 1-i/p$, and $\gamma_i = i/p$~\cite{sack2021}.

However, QAOA is commonly considered a Variational Quantum Circuit (VQC), which means that parameters are optimized so that the output distribution of QAOA has a high probability of sampling the optimal state. For that, classical optimizers are employed, which can make use of gradient estimation techniques like the parameter shift rule~\cite{PhysRevA.98.032309} to accelerate the parameter training. Low-$p$ QAOA circuits are relatively short since $U_M$ can be implemented with just a single layer of $R_X$ gates and $U_C$ with as many $R_{ZZ}$ gates as present in the QUBO. This makes QAOA especially interesting for NISQ-era devices.

\textbf{QAOA Extensions}: As the name discloses, QUBOs cannot handle constraints. Instead, constraints have to be formulated as penalty terms that disallow any constraint-violating states through higher energy contribution. Nevertheless, it is possible to encode constraint-satisfying properties into the QAOA directly in contrast to QA~\cite{hadfield2019}. Some optimization problems require the representation of $k$ categories that require $k$ bits in QUBO, of which only a single bit is set to 1. This is called \emph{one-hot encoding}, e.g. $0100_2$ to represent category 2 of 4. Instead of enforcing this constraint through penalty terms, Ref.~\cite{hadfield2019} develop so-called \emph{single-parity qudit $XY$-mixers} for QAOA, allowing only transitions between valid states. They are defined as follows:
\begin{align}
    U_M(\beta) \rightarrow U_M^{XY}(\beta) = U_\text{last}(\beta)U_\text{odd}(\beta) U_\text{even}(\beta) 
\end{align}
where the layer $U_\text{even}(\beta) = \prod_{i \text{ even}}e^{-i \beta \hat{H}_{XY}(i, i+1)}$ consist of products of non-interacting two-local $XY$-mixers. The $XY$-Hamiltonian is defined as $\hat{H}_{XY}(i,j) = \hat{\sigma}^x_i\hat{\sigma}^x_i + \hat{\sigma}^y_j\hat{\sigma}^y_j$. $U_\text{odd}(\beta)$ is constructed similarly but offset by one qubit to form a brick wall configuration. Lastly, the mixer layer has to be finished by a single $U_\text{last} = e^{-i \hat{H}_{XY}(k, 1)}$ that connects the first and last qubit if $k$ is odd.
The initial state of the $XY$-QAOA has to be prepared as a superposition of valid states, or so-called $\ket{W_k} = k^{-1/2} (\ket{1000} + \ket{0100} + \cdots)$ state, which can be efficiently implemented on quantum hardware~\cite{cruz2019, wang2020}.

Besides one-hot encoding, categories can be represented through \emph{integer encoding} (e.g., $10_2$ representing 2), which requires only a logarithmically growing number of qubits $\lceil \log_2 k \rceil$. Therefore, it is space-efficient, but cost functions are generally more difficult to implement, requiring higher-than-quadratic order qubit interactions, i.e., Higher-Order Binary Optimization (HOBO)~\cite{stein2023, glos2020}. Such interactions can be implemented on universal gate computers, e.g., using CNOT ladders~\cite{glos2020}. If $k$ is no power of two, additional constraints are required to limit the integer space~\cite{glos2020}.

Encoding permutations of $k$ entries requires $k^2$ bits, where each $k$-bit string represents the position of a single entry. The permutation structure has to be enforced using either $2k^2$ one-hot penalty terms or by utilizing a specialized mixer, which only allows for transitions between feasible permutation states. Bärtschi and Eidenbenz~\cite{bartschi2020} developed such a mixer by discovering that a diffusion operation known from the Grover algorithm~\cite{grover1996} and a state preparation circuit for an equal superposition of feasible states suffices for that. They construct the circuit for an superposition of permutations $\ket{S_k} = U_{S_k} \ket{0}$ using at most $O(k^3)$ gates and $O(k^2)$ layers. 
With $U_{S_k}$ and its adjoint, the mixer is defined as follows
$U_M^{S_k}(\beta) = e^{-i \beta \ket{S_k}\bra{S_k}} = U_{S_n} e^{-i \beta \ket{0}\bra{0}}  U^\dagger_{S_n}$, where the central operator is the Grover diffusion.


\section{Benchmarking Process}
\label{sec:methodology}

In this chapter, we outline the crucial elements of a profound benchmarking process and delineate the steps involved as presented in Fig.~\ref{fig:overview}.


\subsection{Selection of Algorithms}




The selection of algorithms is highly dependent on the goal of the benchmarking process. When trying to evaluate the performance of quantum algorithms compared to classical ones in a certain COP, special care should be put into the selection of the best classical algorithms. Quantum algorithms generally require a problem formulation in QUBO form and, therefore, constraints modeled as penalty terms. Classical BNC solvers like Gurobi or CPLEX, on the other hand, do not require unconstrained formulation but instead excel when utilizing MILP formulations~\cite{seker2022}. Thus, the best possible mathematical formulation may differ between classical and quantum methods~\cite{glover2022}. The investigation should always focus on identifying the best formulation for both cases.

Furthermore, classical heuristics should always be considered, as they commonly have a fast runtime to obtain a single sample and may even beat BNC methods~\cite{romero2018, yu2022}.

The selection of quantum algorithms is often problem-dependent, but the QAOA and QA are some of the most frequently used algorithms for quantum optimization~\cite{abbas2023quantum, blekos2023, poggel2023, pelofske2023}. The QAOA, in its more general framework, also allows for constraint-preserving architectures~\cite{hadfield2019, bartschi2020}, which benefits certain COPs.
Reviewing established and promising quantum algorithms, particularly those with theoretical advantages, should be part of a benchmarking methodology.

However, if the goal of a benchmark is to only compare quantum algorithms against each other (e.g., when classical performance is out of scope for quantum algorithms), the selection of classical algorithms is not necessarily required. Nevertheless, mentioning the performance of classical algorithms in solving the investigated problem certainly helps classify the results. For the same reason, it is beneficial to include a naive baseline, such as random sampling~\cite{finzgar2022, pelofske2023}.


\subsection{Suitable Problem Instance Dataset}


Selecting the right dataset involves considering several factors, such as the hardness of the problem instances, its relevance in real-world scenarios, and whether the algorithms being tested are suitable for it. For instance, when simulating quantum algorithms like the QAOA, the number of qubits available imposes strict limitations, which dictates the size of the investigable problems. Similarly, current quantum devices are constrained in the number of qubits and the density of couplings. 

In general, problem instances that mimic real-world structures should be preferred to demonstrate the efficiency of quantum approaches, as this path leads toward quantum utility~\cite{bartz-beielstein2020, abbas2023quantum}. If real-world data is inaccessible, one must rely on randomly generated instances. For that, one should examine phase transitions of the investigated problem beforehand and make sure to sample from the areas in the configuration space where problems can be practically be hard~\cite{dunning2018}. Alternatively, pre-selection of hard instances based on the performance of classical methods can be employed. However, this bears the risk of introducing bias, since only instances that are easy for another solver can be sampled due to no free lunch. In any case, the randomly generated dataset must be generated in a reproducible fashion and should be diverse to prevent results from being non-generalizable across different problem instances.

If meeting the mentioned criteria is infeasible, the research must address these shortcomings transparently.

\subsection{Figures of Merit}

To quantify solver performance, we must compute problem-dependent figures of merit with respect to the outcome. Which solver is considered best heavily depends on the choice of metric, which, in turn, depends on the application's intended purpose. For example, a company may have restricted time to solve a problem, and the quality of the solution may be linked to the costs. Another company may require the optimal solution, and the runtime for finding it then determines the cost.

Except for BNC solvers, heuristics and quantum optimization algorithms produce a distribution of outcomes. Sampling from that distribution $M$-times generates the sample set of bit-strings $X = \{x_i\}$ whose performance we would like to quantify. 

The optimal solution sample will be denoted as $x^*$ and can be found using BNC solvers when the problem size is reasonably small. Should the time to find the optimal solution be infeasible, we need to rely on relative performance. That is, for each problem instance, we extract the overall best-found bit string from all solvers' samples as follows:
\begin{align}
    \hat{x} = \argmin_{x \in \bigcup_s X_s} C(x),
\end{align}
where $X_s$ is the sample set obtained by solver $s$. Whenever the optimal solution is required but unavailable for a metric $\alpha$, we must fall back to $\hat{x}$ and relabel the metric to $\hat{\alpha}$.

Nonetheless, it is important to note that the intrinsic value of a BNC solution is higher than the value of a heuristic solution because a BNC solver can prove optimality.

In the following, we list an extensive set of general metrics that can be employed for benchmarking. Hereby, it is important to discern between holistic methods, i.e., methods that integrate the trade-off between runtime and solution quality, and others. Also, this list is not complete since problem-specific measurements can and should also be considered. 

\subsubsection{Time to Solution} 

One of the main figures of merit is the runtime needed for a solver to find the \emph{optimal} solution. For that, the optimal solution is required beforehand to check whether it has been found. Algorithm runtime without finding the optimal solution cannot be considered Time-To-Solution (TTS).
The wall-clock runtime $t$ of a solver call can generally split into
\begin{align}
    t = t_\text{preprocess} + t_\text{solve} + t_\text{postprocess}.
\end{align}
For BNC solvers, the TTS is precisely the wall-clock time ($\text{TTS} = t_\text{solve}$) without pre- and post-processing overhead. Proof of optimality is not required.

For heuristic multi-sample algorithms, a more sophisticated method for measuring TTS can be applied: TTS can be extrapolated based on the probability of measuring the optimal solution. Sampling from the solver, we can estimate $p^*$, the probability of measuring $x^*$. Then, TTS is defined as follows~\cite{albash2018, lubinski2024optimization, kowalsky2022}
\begin{align}
    \text{TTS}(X) = \frac{t_\text{solve}}{M} \left\lceil\frac{\log(1 - 0.99)}{\log(1 - p^*)}\right\rceil,
\end{align}
where $t_\text{solve} / M$ is the time to obtain a single sample. The remaining term is the number of repetitions required to sample the optimal solution once with a probability of 99\%. In the edge cases $p^* = 1$ and $p^* = 0$, TTS is defined as $t_\text{solve} / M$ and $\infty$, respectively.

The latter case exposes a caveat of this approach: To apply this metric, the solver needs to sample at least a few optimal samples. Otherwise, TTS cannot be computed or is highly inaccurate. When simulating quantum optimization circuits, on the other hand, $p^*$ is exactly determinable through the statevector entries, rendering TTS particularly useful~\cite{PhysRevX.10.021067}. One should note that for classical simulation of quantum circuits, measuring the wall-clock time is not meaningful since simulation overhead scales exponentially with the number of qubits. Instead, estimating the sample time through the number of layers in the quantum circuit is more accurate. A fair comparison of simulated circuits and other solving techniques is otherwise difficult to achieve.

Another challenge for TTS is that the optimal solution has to be accessible prior to computing the metric. Because of this, the applicability to classically challenging problems is limited. As an alternative, \emph{Time-To-Target} (TTT) may be considered, where, instead of the optimal solution, a threshold cost is specified $C_\text{thresh}$ and $p_\text{thresh}$ is the probability of measuring the threshold energy or less~\cite{king2015,lubinski2024optimization}.

Finally, since TTS does not include pre- and post-processing runtime, we also define $\text{TTS}_\text{oh} = \text{TTS} + t_\text{preprocess} + t_\text{postprocess}$ to include the overhead time into the metric.

\subsubsection{Best Solution Found}
In real-world applications, large optimization problems need to be solved within certain time bounds. For example, a vehicle routing problem will be optimized overnight when no deliveries happen. But, it must be operational as soon as the first vans are loaded with parcels. Therefore, in such a use case, we are not interested in which solver can find the optimal solution fastest but in which solver produces the Best Solution Found (BSF) within a given time frame.

For comparability of found solutions, we define the relative cost of the BSF in comparison to the optimal solution~\cite{dunning2018, Oshiyama2022}
\begin{align}
    c(X) = \frac{\min_{x \in X} C(x)}{C(x^*)}.
\end{align}
The advantage of this metric lies in its simplicity and universality. It is straightforward to calculate and applicable in nearly all scenarios. Furthermore, it is also applicable even if the optimal solution is unknown through the usage of $\hat{c}$. Sometimes, it is beneficial to use the relative error (or optimality gap) $|1 - c|$ instead~\cite{seker2022}.

Ensuring solver comparability is crucial because solution quality depends on the resources attributed to a strategy. Ideally, all solvers would utilize the exact same resources, but this is generally challenging to achieve. A common approach is to apply the same time constraints to all algorithms~\cite{Oshiyama2022}, thereby aligning the evaluation criteria for both classical and quantum algorithms. Heuristic solvers, like quantum algorithms, must be repeated until the time limit has been reached. However, BNC algorithms can already prove the optimality of a solution and terminate before all planned resources are used, which can be a big advantage in operation.

This metric can be prone to statistical fluctuations, especially if the sample size is small. As described earlier, the samples follow a distribution, and the best sample is located at the edge of the distribution. Thus, large sample sizes are generally required.

\subsubsection{Fraction of overall best solutions}
To directly compare solvers over an instance dataset, it can be beneficial to count how often a solver finds the overall best solution. For that, we define the Fraction of Overall Best solution found (FOB), defined over all samples of all instances $\mathcal{I}$, similar to~\cite{dunning2018, Oshiyama2022},
\begin{align}
    \text{FOB}(\{X_i\}) = \frac{|\{i\,|\hat{c}(X_i) = 1, \forall i \in \mathcal{I}\}|}{|\mathcal{I}|},
\end{align}
where $X_i$ the samples obtained for a single instance $i$.

\subsubsection{Pareto Front}
If the trade-off between runtime and solution quality cannot be eliminated by fixing one of the two variables (e.g., TTS: solution quality is fixed to the optimal solution, BSF: runtime resources are fixed among solvers), comparability can still be achieved by analyzing the Pareto front of runtime, solution quality, and optionally more metrics~\cite{sharma2022}.

\subsubsection{Approximation Ratio}
A metric that considers the whole sample set is the Approximation Ratio (AR) based on the expectation value $\langle C \rangle_X = \frac{1}{M}\sum_{x \in X} C(x)$. Single outliers do not influence the result as much as the BSF metric. AR is defined as the normalized expectation value~\cite{sack2021, PhysRevX.10.021067, blekos2023}
\begin{align}
    r(X) = \frac{\langle C \rangle_X}{C(x^*)}.
\end{align}
Sometimes AR is defined to lie between $0$ and $1$ by using the worst solution $\max_x C(x)$ as well~\cite{abbas2023quantum}.


A drawback of AR is that different algorithms need different time intervals to create one sample, therefore it is not holistic. When total runtime is increased, AR becomes more accurate. So, fixing the runtime for all solvers leads to comparing ARs obtained from different sample sizes, which is potentially inaccurate (especially if sample sizes are small, like for BNC solvers).

However, AR is especially useful for comparing different settings of the same heuristic or quantum solver, i.e., measuring how the whole sample set improves when certain hyperparameters are tuned. In practice, AR is commonly used as the metric used to optimize the parameters in a VQC~\cite{blekos2023}.

\subsubsection{Feasibility Ratio}
Quantum algorithms often require reformulating constraint optimization problems into unconstrained ones. Consequently, infeasible solutions can be sampled when faced with an unconstrained formulation. The feasibility ratio~\cite{finzgar2022}, defined as follows
\begin{align}
    f(X) = \frac{1}{M} \sum_{x \in X} \mathbb{1}_\mathcal{F}(x),
\end{align}
measures how many samples do not violate the constraints. Here, $\mathbb{1}_\mathcal{F}$ denotes the characteristic function of the feasible set $\mathcal{F}\subseteq\mathcal{X}$. BNC solvers, as well as heuristics that only search the feasible space, will have a default feasibility ratio of 1. The feasibility becomes vital when comparing different formulations or methods of mapping a problem onto quantum hardware~\cite{glos2020}.

\subsection{Scaling analysis}

Besides comparing absolute metric values, it is advantageous to consider the scaling of a metric with respect to the problem size. In some cases, certain methods bear an overhead, but the scaling of the metrics suggests that an inflection point at larger problem instances may happen~\cite{finzgar2022}. The different analyses of the scaling behavior alleviate comparability issues when the hardware is not directly comparable, e.g., QPU, CPU, GPU, or cloud services.

Furthermore, it is vital to consider scaling quantum resources because the number of available qubits is vastly limited. Similarly, the depth of the quantum circuits also plays a role, as NISQ-era qubits are highly error-prone~\cite{pelofske2023}. 

\subsection{Hyperparameter Optimization}

Hyperparameter optimization is a crucial process to unlock the full potential of algorithms. This optimization can be approached in several ways: iterative tuning of each parameter, selective optimization of key parameters, exhaustive grid searches across all parameters~\cite{lerman1980}, or employing random parameter selection~\cite{bergstra2015}. More advanced methods, such as genetic algorithms~\cite{erden2023} or Bayesian methods~\cite{akiba2019}, offer further sophisticated optimization strategies.

It is essential to perform hyperparameter optimization tailored to a specific metric and dataset since outcomes can significantly vary with changes in problem size, type, or figure of merit. Equally, ensuring consistent effort in optimizing hyperparameters across all algorithms is vital. Differential levels of optimization effort can markedly affect the comparative results; extensive optimization for one algorithm may enhance its performance significantly, whereas another might not reach its potential due to a lack of similar optimization.


\subsection{VQC Parameter Optimization}

Optimization algorithms based on VQCs suffer from the learning overhead caused by training each of the circuit's parameters to a specific problem instance. Therefore, one has to attribute the full classical optimization process to the quantum optimization algorithm's runtime.

However, parameter optimization can also be seen as a generalization task similar to machine learning~\cite{montanez-barrera2024}. In that sense, the parameters that are searched for need to return the best solution quality averaged over a set of input problems. Reusing the parameters for benchmarking allows us to consider only the execution time of the single optimization circuit. Much like in machine learning, it is also important to note that the instance set on which the parameters have been trained is not the same as the benchmark set to minimize bias.

In the case of QAOA, decaying $\beta_i$s and growing $\gamma_i$s are hinted by the connection to QA~\cite{sack2021}. To regularize this pattern for the parameters and avoid overfitting, we propose the usage of a lesser parametrized generating function $g$ for the QAOA parameters,
\begin{align}
    \beta_i = g(\theta_\beta, i / p) \qquad \gamma_i = g(\theta_\gamma, i / p)
\end{align}
where the parameters $\theta_\beta$, $\theta_\gamma$ are subsequently optimized. For instance, a low degree polynomial may be used $g(\theta, x) = \theta_0 + \theta_1 x + \theta_2 x^2 + \cdots$. This is similar to the Fourier parameter method of Ref.~\cite{PhysRevX.10.021067}.

\section{Examples}
\label{sec:examples}
In this section, we apply our guidelines to two MC scenarios and one TSP scenario. Through these examples, we aim to demonstrate how our approach facilitates decision-making processes amidst multifaceted considerations. The demonstration underscores the applicability and significance of these guidelines in various real-world scenarios.

All numerical experiments are conducted on an AMD Ryzen Threadripper PRO 5965WX (single-core). For QA, we utilize D-Wave Advantage 5.4 (DW)~\cite{mcgeoch}.

\subsection{Max-Cut}
 
\begin{table}
    \centering
    \caption{MaxCut instance scenarios}
    \renewcommand{\arraystretch}{2.5}
    \def\wd{2.35cm}
    \begin{tabular}{cccccc}
        \# & Nodes & Edges & Instances & Solvers & FoM \\
        \hline
        1 & 10--20 & 20--143 & \begin{minipage}{1.45cm}\centering 3-Regular, \erdosrenyi{}\end{minipage} & \begin{minipage}{\wd}\centering Gurobi, CPLEX, SA, TS, GW, Greedy, DW, QAOA\end{minipage} & TTS\\
        2 & 99--298 & 110--2000 & MQLib & \begin{minipage}{\wd}\centering Gurobi, CPLEX, SA, TS, Greedy, DW\end{minipage} & BSF 
    \end{tabular}
    \label{tab:mc-scenarios}
\end{table}

Max-Cut (MC) is an NP-hard optimization problem that is stated as follows: Given an undirected graph with weighted edges $G = (V, E, w)$, find a partition of the vertices into two sets such that the total weight of edges crossing the sets is maximized. Using binary variables $x_i \in \{0, 1\}^{n}$, with $n = |V|$, we can formulate MC as QUBO
\begin{align}\label{eq:mc-formulation}
    C(x) = \sum_{(i, j) \in E} (2x_i x_j - x_i - x_j) w_{i,j},
\end{align}
i.e., giving a negative contribution if $x_i \neq x_j$ (assigned to different sets).

MC is a fundamental NP-hard problem that occurs in many theoretic investigations, as well as a few practical use cases, such as Very Large Scale Integrated Circuit (VLSI) problems~\cite{dunning2018}.

In the following, we consider two MC benchmarking scenarios, overviewed in Tab.~\ref{tab:mc-scenarios}.

\subsubsection{Scenario 1}

In the first scenario, we aim to benchmark the performance of quantum algorithms and classical heuristics on small MC instances. We especially want to consider an idealized simulated QAOA algorithm, which limits the problem size, as simulation becomes exponentially expensive with the number of involved qubits.

\paragraph{Algorithms} Besides QAOA at different depths $p$, we consider DW, SA, TS, Gurobi, and CPLEX. Both Gurobi and CPLEX use the QUBO MC formulation given in Eq.~\ref{eq:mc-formulation}. In addition, an ILP formulation~\cite{barahona1986} has been assessed but showed inferior runtime.

Furthermore, we investigate two MC heuristics. A naive Local Search (LS) strategy that starts out with a random assignment and then iterates over all nodes and improves the cut size when it is possible by moving the current node into the other partition. This iteration is repeated until the cut size cannot be improved anymore. The second is based on the Goemans-Williamson (GW)~\cite{goemans1995} Semi-Definite Programming (SDP) relaxation of the MC problem. The convex SDP problem is solved using \texttt{cvxpy}~\cite{diamond2016cvxpy}, revealing the quadratic matrix $x_i x_j = X_{i,j}$. By randomly picking a hyperplane that separates the relaxed $x$ vectors into two categories, we can sample cuts that approximate the problem with AR $r \approx 0.878$~\cite{goemans1995}.

\paragraph{Dataset} To be able to simulate QAOA efficiently, the benchmark is limited to MC problems consisting of 10--20 nodes. Real-world problem instances are not common at this size. Therefore, we randomly generate four graph types: \erdosrenyi{}~\cite{erdos2022} graphs with 25\%, 50\%, and 75\% connectivity and 3-regular graphs (meaning every node is connected to precisely three other nodes). Our dataset consists of 50 instances per graph type and problem size, i.e., a total of 1200 graphs.

\paragraph{Figure of Merit} We choose TTS as the central metric since the exact solution is easily accessible $p^*$ can be well estimated for all instances. As a secondary objective, we consider the approximation ratio that helps to estimate the distribution of samples for heuristic solvers.

\paragraph{Hyperparameter tuning} Since the instances are random, we generate a second dataset solely used for hyperparameter optimization to avoid overfitting. For SA, we observed that just a small number of Monte Carlo sweeps (20) is required to minimize TTS on average. The geometric annealing schedule performed best. TS did not significantly respond to changing tenure and restart parameters, therefore we keep them default. DW performed well with default settings. To gather sufficient statistics for $p^*$, we perform 1000 runs for each heuristic. CPLEX and Gurobi are considered out-of-the-box solvers without hyperparameters. LS and GW don't have parameters.

For the QAOA, we use a degree-4 polynomial as a generator function and adjust the hyperparameters for each QAOA-depth $p = 2,4,8,16, \text{and } 32$ separately by minimizing the mean over the approximation ratios of training instances. For minimization, we utilize the popular gradient-based L-BFGS optimization algorithm~\cite{nocedal1980}.

\begin{figure}
    \centering
    \includegraphics[width=0.9\columnwidth]{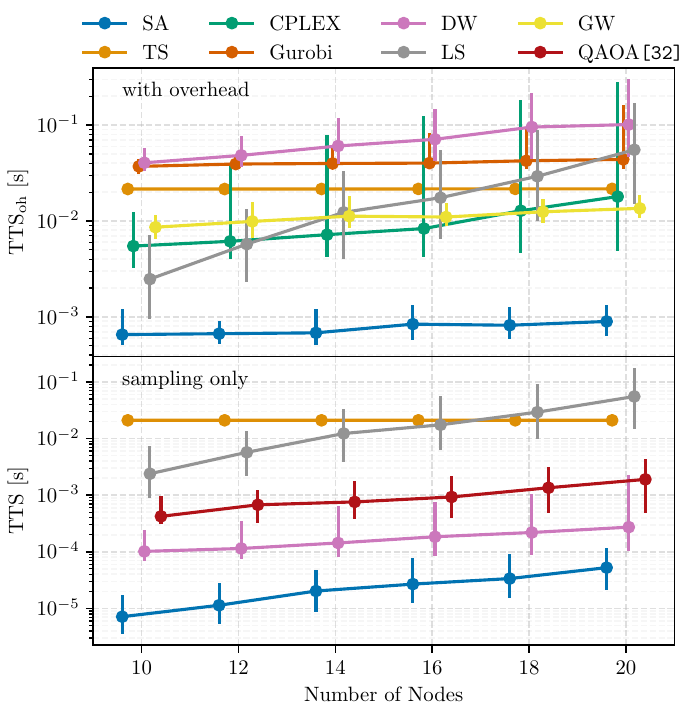}
    \caption{TTS with and without overhead for different Max-Cut solvers on problems. Error bars indicate the 75\% percentile interval. QAOA is not present in the TTS$_\text{oh}$ because of the simulation; a CNOT layer is estimated to take 1$\upmu\mathrm{s}$. For the sampling only TTS, we only consider heuristics.}
    \label{fig:mc-tts}
\end{figure}
\begin{figure}
    \centering
    \includegraphics[width=0.9\columnwidth]{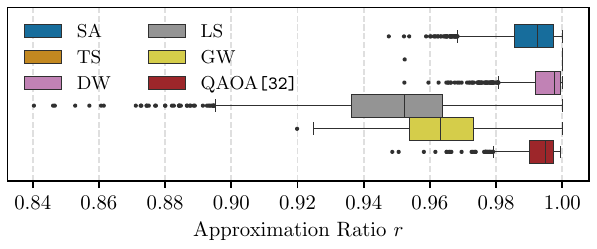}
    \caption{AR from all sizes for heuristic MC solvers. TS finds the optimal solution in almost every sample. It is important to note that the runtimes for estimating AR are vastly different.}
    \label{fig:mc-approx-ratio}
\end{figure}
\begin{figure}
    \centering
    \includegraphics[width=0.9\columnwidth]{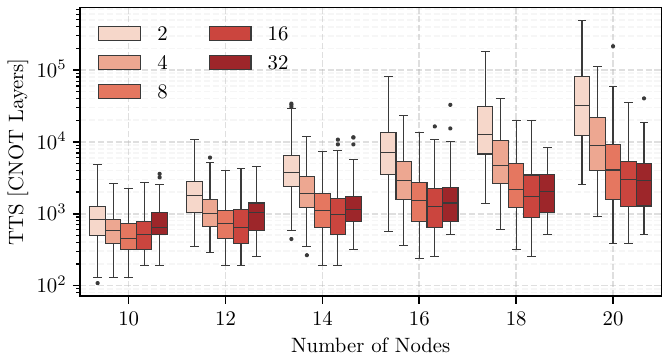}
    \caption{Max-Cut QAOA TTS in terms of CNOT layers for different QAOA rounds $p = 2,\dots,32$. A U-shape is observable for every problem size, indicating an optimal QAOA depth.}
    \label{fig:mc-qaoa}
\end{figure}
\paragraph{Results} Fig.~\ref{fig:mc-tts} shows the TTS$_\text{oh}$ and TTS of the experiments with the different solvers. As apparent, SA performs best with the lowest TTS (with and without overhead).

We must consider both TTS with and without overhead here since some solvers have significant overhead. DW requires time-consuming embedding before any sampling can be achieved. As for GW, the main runtime comes from the SDP relaxation, which is not part of the sampling process. The BNB solvers Gurobi and CPLEX exhibit relatively high runtime, which can be explained by the large overhead of the algorithms, being too complex for the simple problems at hand.
TS has almost constant TTS because almost every sample is optimal (see Fig.~\ref{fig:mc-approx-ratio}).
LS shows the steepest runtime increase of all considered solvers.

The AR comparison of all instances, Fig.~\ref{fig:mc-approx-ratio}, shows that GW and LS sample the most widespread results. SA, DW, and QAOA, on the other hand, show similar ARs. Here, it is again important to note, that a higher number of MC sweeps would improve the AR of SA but deteriorate TTS as a trade-off.

Finally, Fig.~\ref{fig:mc-qaoa} shows the QAOA's TTS regarding CNOT layers required to find the optimal solution. A single CNOT layer can feature maximally $n/2$ non-overlapping CNOT gates for $n$ qubits. The number is estimated by the amount of $R_Z$ gates required to encode the QUBO in the quantum circuit, which can be computed by finding the edge-coloring of the MC graph~\cite{ostrowski2020}.

As the QAOA depth increases, the solution quality (AR) improves. However, the increased improvement of sampling the optimal solution might be too little to justify the expanded circuit depth. This trade-off can be well observed in Fig.~\ref{fig:mc-qaoa}, especially in smaller instances. For larger instances, deeper circuits always seemed to deliver the optimal solution fastest.

For the comparison of TTS in Fig.~\ref{fig:mc-tts}, one CNOT Layer is estimated with 1\,$\upmu\mathrm{s}$\footnote{Current hardware (\texttt{ibm\_brisbane}) has a two-qubit gate time of $660\,\mathrm{ns}$.}
The sampling TTS of DW and QAOA\texttt{[32]} is very similar in scaling behavior. The examined scaling behavior of the quantum solvers (DW and QAOA) is shallower than both SA and LS.

\paragraph{Discussion}
As apparent from the results, SA performs best over all instances. DW obtains competitive TTS ($\approx 10 \times$ slower than SA, but the overall TTS$_\text{oh}$ is considerably worse due to embedding. QAOA's TTS scales are similar to those of DW, but the absolute values are difficult to compare because of the CNOT layer execution time estimation. The other heuristics and BNC solvers perform worse overall.
\subsubsection{Scenario 2}

The second scenario assesses the effectiveness of solving larger-sized MC instances within a time limit of 10 seconds.

\paragraph{Algorithms} We consider the same solvers as used in the first scenario, except for the QAOA (due to problem size) and GW, since it demonstrated difficulties in solving the relaxed problem within a feasible timeframe.

\paragraph{Dataset} We examine MC problems consisting of 99--298 nodes with 110--2000 edges from the MQLib~\cite{dunning2018}. This selection contains both real-world (VLSI) and randomly generated problems, resulting in a highly diversified dataset. We limit the maximum number of edges to ensure most instances from the dataset are embeddable on the DW hardware.
In total, the dataset comprises 371 instances.

\paragraph{Figure of Merit} The investigated metric here is BSF since the $p^*$ cannot be estimated accurately anymore. The relative cost is calculated as the cut size relative to the best cut size found among all algorithms $\hat{c}$. We compare the relative error $1 - \hat{c}$ in our experiments. Furthermore, we also consider FOB. Each solver has a time limit of 10\,s per instance.

\paragraph{Hyperparameter tuning} The setting for the heuristic solvers studied is as follows~\cite{atkinson1992}: The sampling process of SA, TS, and DW is repeated until the runtime limit is reached. The parameterization remains fixed throughout all repeats. All instance-specific preprocessing, e.g., embedding, is cached in the initial iteration and reused subsequently. 
SA and TS are sampled once per iteration, for DW batches of 1000 reads are processed, as the network overhead is significantly larger than the sampling itself. The batching has no effect on the solution quality and merely reduces overhead.

Given the random and real-world nature of the instances, we optimize the hyperparameters on a subset of 50 randomly selected instances using grid search. For SA, we observed that a large number of Monte Carlo sweeps ($2 \times 10^4$)---albeit limiting the number of repetitions within the time frame---with an otherwise default parameterization yields the best results. Akin to the first scenario, TS did not respond to tuning. For DW, the most effective configuration involved utilizing the default annealing time of 20\,$\upmu \mathrm{s}$ in conjunction with a chain strength factor~\cite{willsch2022} of 0.2. LS, Gurobi, and CPLEX do not entail hyperparameters.

\begin{figure}
\centering
\includegraphics[width=0.9\linewidth]{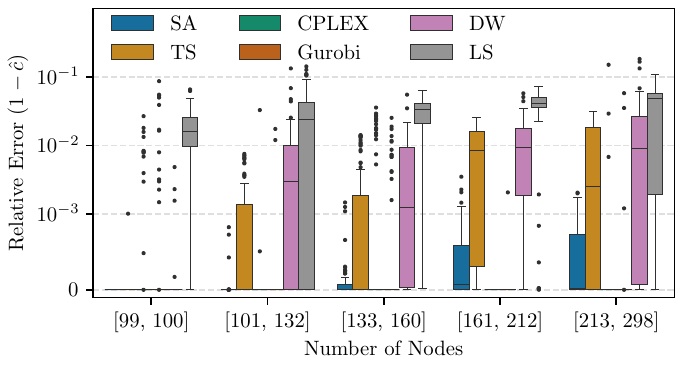}
\caption{Relative error in cut size within a group. The time limit is set to 10\,s.}
\label{fig:maxcutmedrelcut}
\end{figure}

\begin{figure}
\centering
\includegraphics[width=0.9\linewidth]{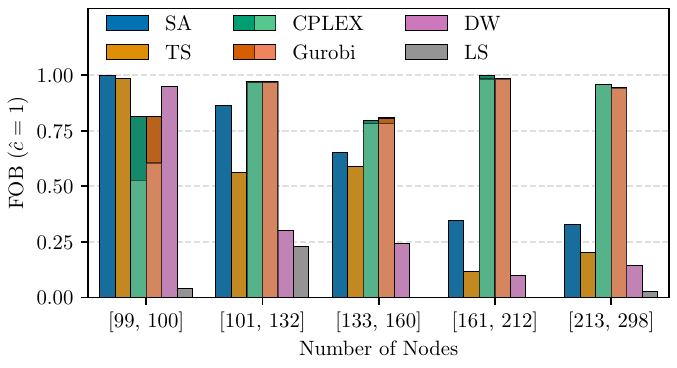}
\caption{FOB within a group. The lighter shading for Gurobi and CPLEX indicates that the solver proofed optimality.}
\label{fig:maxcutmedisopt}
\end{figure}

\paragraph{Results} An embedding on the DW hardware was found for 351 instances. Subsequent analyses will concentrate on these specific instances for comparative purposes.
The results are organized into groups, ensuring that each group comprises roughly the same number of problem instances.
Fig.~\ref{fig:maxcutmedrelcut} illustrates that Gurobi and CPLEX achieve the lowest relative error in cut size ($\leq 10^{-4}$) across all groups, except the first group (99--100 nodes) comprising the most densely connected graphs within the dataset, where Gurobi and CPLEX have significantly more outliers with a much higher relative error compared to SA, TS, and DW.
Following closely is SA; however, the results indicate a progressive increase in error as the number of nodes increases. DW and TS exhibit comparable results across all groups but are clearly outperformed by SA, CPLEX, and Gurobi.

The evidence from Fig.~\ref{fig:maxcutmedisopt}, depicting the FOB, strongly supports these findings.
For the first group, Gurobi and CPLEX are clearly outperformed by SA, TS, and DW. While SA achieves a FOB of 1.0, closely followed by TS ($\approx 0.99$) and DW ($\approx 0.95$), Gurobi and CPLEX only achieve a FOB of $\approx 0.82$. This insight, combined with the results from Fig.~\ref{fig:maxcutmedrelcut}, signals a diminishing efficacy of both CPLEX and Gurobi in handling graphs of higher density.
Apart from the first group, Gurobi, and CPLEX---closely followed by SA up to 160 nodes---attain the overall best-found solution significantly more often than the other algorithms while consistently verifying that the optimal solution was found. 
Across all instances, Gurobi slightly outperforms CPLEX, as it proves to find the optimal solution for 2\% more instances.

The LS algorithm exhibits notably inferior performance compared to all other algorithms, characterized by the highest relative error and the lowest FOB across all instances (Fig.~\ref{fig:maxcutmedisopt}).

\paragraph{Discussion} The heuristics SA, TS and DW perform significantly better on the highly dense graphs within the dataset. Notably, DW achieves a high FOB and maintains a low relative error, especially when compared to the classical state-of-the-art solvers Gurobi and CPLEX. However, for larger and sparser graphs outside this subset, the BNC solvers consistently outperform the others, exhibiting the lowest relative error and achieving the highest FOB.

\subsection{Travelling Salesperson Problem}

The Travelling Salesperson Problem (TSP) is one of the most renowned COPs structured as follows: Given a fully connected weighted graph with $k + 1$ nodes (or locations), find the shortest cycle that visits every node once. Therefore, we can fix the first node as our starting location, leaving $k$ nodes unassigned. In a mathematical sense, TSP can be expressed as follows
\begin{align}
    \argmin_{\pi \in S_k} \left(\sum_{\ell = 1}^{k - 1} d_{\pi_\ell, \pi_{\ell + 1}} +  d_{0, \pi_1} + d_{\pi_k, 0}\right),
\end{align}
where $d_{i,j}$ the distance between location $i$ and $j$ and $\pi \in S_k$ is a permutation in the symmetric group.


\begin{table}
\centering
    \caption{Estimation of CNOT circuit layers required for different QAOA implementations for an $k + 1$ location TSP problem. }
    \def\arraystretch{1.3}
    \begin{tabular}{c|cccc}
        Metric & QUBO & HOBO & XY-Mixer & Perm \\
        \hline 
        State Prep. & $-$ & $-$  & $4\lceil \log_2 k \rceil$ & $O(k^2)$ \\
        Cost Layer & $8k$ & $2 k^3$ & $6k$ & $4k$ \\
        Mixer Layer & $-$ & $-$ & $8 \, (+4)$ & $O(k^2)$ \\
        \hline
        Qubit Count & $k^2$ & $k\lceil\log_2 k\rceil$ & $k^2$ & $k^2$\\
        Search Space & $2^{k^2}$ & $2^{k \lceil \log_2 k \rceil}$ & $k^k$ & $k!$
    \end{tabular}
    \label{tab:tsp-qaoa-specs}
\end{table}

\begin{figure*}
\centering
\subfloat[TTS in terms of CNOT layers with respect to TSP problem size\label{fig:tsp-tts}]{\includegraphics[scale=0.72]{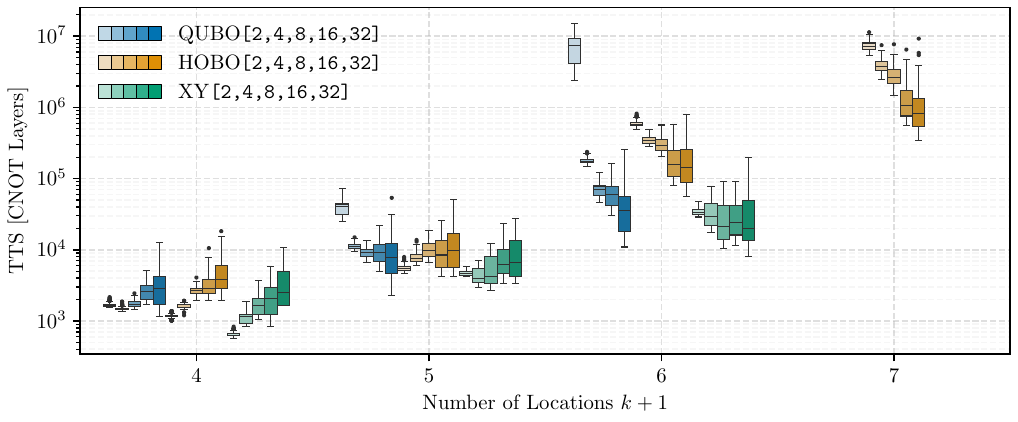}}%
\hfill%
\subfloat[TSP solution quality error \eqref{eq:tsp-sol-error}\label{fig:tsp-sol-quality}]{\includegraphics[scale=0.72]{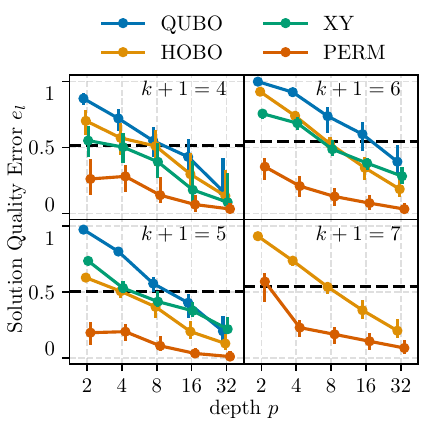}}
\caption{%
Results of the QAOA-only benchmark on the TSP. 
TTS is shown in pane (a), and the relative error is shown in pane (b). Error bars indicate the 75\% percentile interval. The dashed line indicates random sampling from the feasible space.
}\label{fig:tsp-results}
\end{figure*}

\paragraph{Algorithms and Formulations} TSP can be formulated as an ILP with lazy constraints\footnote{\url{https://www.gurobi.com/jupyter_models/traveling-salesman/}}, which is a highly efficient implementation. In fact, solving a 100-node problem takes less than a second on a laptop, which is out of reach for any current quantum solver since the QUBO formulation already requires $k^2 = 10\,000$ qubits.

Therefore, we limit the TSP benchmark to only QAOA, comparing the performance of different formulations and implementations. The QUBO with the one-hot encoded binary variables $x \in \{0,1\}^{k\times k}$ is given by~\cite{lucas2014}:
\begin{gather}
    C(x) = A \sum_{i=1}^{k}d_{0,i} (x_{i,1} + x_{i,k}) + A \sum_{i, j = 1}^k d_{i,j} \sum_{t=1}^{k-1} x_{i,t} x_{j,t + 1} \nonumber\\
    + B \sum_t \left(1 - \sum_i x_{i,t}\right)^2
    + B \sum_i \left(1 - \sum_t x_{i,t}\right)^2,
\end{gather}
where the $B > A \max_{i,j} d_{i,j}$ are penalty factors that ensure the one-hot constraints, penalized by the $B$-terms, are satisfied~\cite{lucas2014}. We set $B = 1$ and $A = \frac{1}{1 + \max d_{ij}}$.

As discussed in Sec.~\ref{sec:background}, various methods are available to handle one-hot constraints or permutations with QAOA. We, therefore, strive to compare QAOA with QUBO~\cite{lucas2014}, HOBO~\cite{glos2020}, $XY$-Mixer~\cite{hadfield2019}, and direct Grover permutation Mixer~\cite{bartschi2020}. We simulate QAOA up to $k+1=6$, i.e., 25 qubits in the QUBO and $XY$ case. HOBO is more space efficient and can be simulated for larger problem instances. Simulation of the Grover diffusion mixer is straightforward and only requires $k!$ statevector entries to be stored. An overview of the qubit requirements and search space sizes can be found in Tab.~\ref{tab:tsp-qaoa-specs}.

An alternative method for solving the TSP with QAOA through the ILP with lazy constraints exists~\cite{montanez-barrera2023}, but is not further discussed because of its iterative nature.

\paragraph{Dataset} The qubit scaling limits the problem size extensively. Real-world problems hardly exist for this problem scale. Therefore, we fall back to randomly generated instances. Ref.~\cite{schawe2016} proposes a parameterized TSP instance generator that undergoes a phase transition from easy to complex problems. Essentially, it places locations around the circumference of a circle and then offsets them by a random distance (scaled with $\sigma$) in any direction. Additionally, we sample locations randomly in a two-dimensional plane.

We use a naive heuristic that successively chooses the closest unvisited node as the next location to filter the randomly generated instances. We only consider problems in which the heuristic fails to find the optimal solution independent of the starting location.

In total, our dataset consists of 150 circular instances from~\cite{schawe2016} with the parameter $\sigma = 0.6, 1.0, 1.4$ and 50 random instances per problem size.

\paragraph{Figure of Merit} We chose TTS regarding CNOT layers since it also incorporates different circuit complexities into the comparison; see Tab.~\ref{tab:tsp-qaoa-specs}. The permutation-based ansatz has $O(k^2)$ circuit layers per mixer application. However, the precise CNOT counts are inaccessible without full implementation of the circuit, but a large overhead is expected since $U_{S_k}$ is non-trivial~\cite{bartschi2020}.

Additionally, we consider an adapted relative error as a combination of average path length and feasibility ratio as solution quality, defined as
\begin{align}\label{eq:tsp-sol-error}
    e_l(X) = \frac{1}{l_\text{worst} - l^*} \frac{1}{M} \sum_{x\in X} \begin{cases}
    l(x) - l^* & \text{if $x$ feasible} \\
    l_\text{worst} - l^* & \text{else},
    \end{cases}
\end{align}
where $l^*$ and $l_\text{worst}$ are the best and worst solutions found with Gurobi, and $l(x)$ returns the path length of a valid solution $x$.

\paragraph{Hyperparameter Tuning} Similar to the first MC scenario, we use a degree-4 polynomial as a generator function for the QAOA parameters on a separate training dataset. Parameters for each QAOA method are trained separately using gradient-based L-BFGS~\cite{nocedal1980} optimization. Different QAOA rounds $p$ are trained separately.

\paragraph{Results} 
The TTS for the QUBO, HOBO, and $XY$ methods, depicted in Fig.~\ref{fig:tsp-tts}, reveals that the $XY$ method performs best, especially considering low QAOA rounds $p = 2, 4$. At $p > 8$, the plain QUBO and $XY$ methods are comparable in performance. It is apparent that in small instances, the added complexity through higher depths does not offset the increased solution quality, i.e., low depths solve the problem faster throughout all methods. Nevertheless, this behavior inverts when more nodes are considered. The HOBO formulation is comparable up until $k + 1= 5$; after that, $\lceil \log_2 k \rceil = 3$, and therefore, the number of terms increases the circuit depth per QAOA layer drastically.

Considering the combined error in Fig.~\ref{fig:tsp-sol-quality}, we observe that the permutation-based ansatz returns the best results overall, as expected since no infeasible solutions are possible. Nevertheless, TTS comparison to the other approaches is infeasible due to the intricate (high overhead) circuits. The solution quality here shows a comparable (almost better) performance of the HOBO approach compared to the $XY$ mixer algorithm. However, this is a pitfall since it does not consider the overall runtime of the circuit, which is inadvertently larger (see Tab.~\ref{tab:tsp-qaoa-specs}). This, again, proves why holistic metrics like TTS are useful and required for benchmarking.

\paragraph{Discussion} The HOBO QAOA is only useful when the number of qubits is limited since TTS increases massively due to the circuit depth of the higher-order terms. Small-depth QUBO QAOA performs worst but increases considerably with depth $p$. The $XY$-Mixer method shows the best overall TTS performance. The Grover mixer permutation-only formulation produces the best results based on solution quality. However, it is not clear how it compares in TTS with the other methods.

\section{Conclusion}
\label{sec:conclusion}

This paper identified a set of steps and guidelines for the goal of ensuring fair benchmarks of quantum solvers against classical methods, summarized as follows.

We highlight that a meaningful comparison must include state-of-the-art algorithms, both classical (BNC, heuristic) and quantum methods, tested on a selection of problem instances derived from real-world problems or randomly generated data, with some proof of classical hardness.


The selection of the figure of merit is central to every benchmark. Essentially, a meaningful metric should include the trade-off between runtime and solution quality. We feature TTS and BSF within time constraints as two examples.

The hyperparameters of each solution algorithm need to be tuned with equal care to remove initial bias. 
Parameter learning of VQCs can be considered a hyperparameter optimization step, where parameters are learned for a class of problem instances.

We verified the proposed schema on two Max-Cut problem scenarios using TTS and BSF on the smaller and larger instances, respectively. We have found that SA performs best on the small random instance, while the BNC solver outperforms on the larger problem instances with real-world data.

For TSP, we compared different QAOA implementations and found that the low circuit complexity $XY$ approach works best with regard to TTS.

Finally, it remains to emphasize that the proposed steps are not complete but offer a selection of reasonable actions for fair benchmarks. Every considered problem might add additional metrics or considerations that this general framework cannot encompass.


\bibliographystyle{IEEEtranDoi}  
\bibliography{bstcontrol,references} 

\end{document}